\documentclass[aps,prl,twocolumn,showpacs,groupedaddress,superscriptaddress]{revtex4-1}

\usepackage{graphicx}
\usepackage{dcolumn}
\usepackage{bm}
\usepackage{color}%
\usepackage{amsmath}
\usepackage{amsfonts}
\usepackage{array}
\usepackage{multirow}

\begin{document}

\title{Electronic properties of bilayer graphenes strongly coupled to interlayer
stacking and an external electric field}

\author{Changwon Park}
\affiliation{Center for Nanophase Materials Sciences, Oak Ridge National
Laboratory, Oak Ridge, Tennessee 37831, United States}

\author{Junga Ryou}
\affiliation{Department of Physics and Graphene Research Institute, Sejong
University, Seoul 143-747, Korea} \author{Suklyun Hong}
\affiliation{Department of Physics and Graphene Research Institute, Sejong
University, Seoul 143-747, Korea} \author{Bobby Sumpter}
\affiliation{Center for Nanophase Materials Sciences, Oak Ridge National
Laboratory, Oak Ridge, Tennessee 37831, United States}

\author{Gunn Kim}
\email{Corresponding Author: gunnkim@sejong.ac.kr}
\affiliation{Department of Physics and Graphene Research Institute, Sejong
University, Seoul 143-747, Korea}

\author{Mina Yoon}
\email{Corresponding Author: myoon@ornl.gov}
\altaffiliation{Notice: This manuscript has been authored by UT-Battelle, LLC, under Contract No. DE-AC0500OR22725 with the U.S. Department of Energy. The United States Government retains and the publisher, by accepting the article for publication, acknowledges that the United States Government retains a non-exclusive, paid-up, irrevocable, world-wide license to publish or reproduce the published form of this manuscript, or allow others to do so, for the United States Government purposes.}
\affiliation{Center for Nanophase Materials Sciences, Oak Ridge National
Laboratory, Oak Ridge, Tennessee 37831, United States}

\date{ \today }

\begin{abstract}

Bilayer graphene (BLG) with a tunable bandgap appears interesting as an
alternative to graphene for practical applications, thus
{\color{black}its transport properties are being actively pursued.} Using density
functional theory and perturbation analysis, we investigated, under an external electric field, the electronic
properties of BLGs {\color{black}in various stackings
relevant to recently observed complex structures.
We established the first phase diagram summarizing the
stacking-dependent gap openings of BLGs for a given field.  We further
identified high-density midgap states, localized on grain
boundaries, even under a strong field,  which can considerably
reduce overall transport gap.}
\end{abstract}


\maketitle

The discovery of graphene has opened new avenues for studying the role of
dimensionality on the fundamental properties of materials ~\cite{Novoselov}.
Although graphene shows excellent electrical properties ~\cite{Bolotin}, the
zero bandgap of graphene limits its practical application as an electronic
device. On the other hand, gap opening is possible in BLG, thus making it a very
promising material that overcomes graphene's key limitation while retaining many
of its interesting properties.  For example, massive Dirac fermions in BLG
exhibit a bandgap tunable by applying a transverse electric field
(E-field)~\cite{MinMcCann2}; this has been demonstrated by
optical~\cite{OhtaMakZhang} and electrical transport measurements using dual-gated
devices~\cite{Tay,OostingaVelascoVarlet}.  However, these
measurements leave a couple of unsolved problems: 1) the origin of unexpectedly
small transport gaps that are two orders of magnitude smaller than optical
gaps~\cite{Tay} and 2) the origin of anomalous low-temperature ($<$~2 K)
transport behaviors dominated by hopping between localized midgap states,
presumably induced from disorders or defects ~\cite{Tay,Tanabe}.

Recent experiments have revealed complex configurations in BLG, including
various stacking domains induced by rotational faults and soliton
formation~\cite{Lin,Alden,Hibino}.  While AB stacking is energetically most
favorable, the non-AB-stacking region can be stabilized by a minute
twist~\cite{Kim} and the stacking boundary~\cite{Lin}.  The local
{\color{black}stacking} configuration is strongly coupled to its electronic
structure and its response to an external E-field. Therefore, it is critically
important, fundamentally and practically, to understand the observed complex
stackings and their impact on the overall electronic properties.

In this letter, using the framework of an effective Hamiltonian based on density
functional theory (DFT) and perturbation theory, we analyze gap-opening
properties of BLGs near the high-symmetry stackings (AA, AA$^\prime$, and AB),
under an applied E-field.  We establish a phase diagram for the
stacking-dependent gap openings, and further identify grain boundaries
containing non-AB stackings as a source for high-density midgap states even
under a strong E-field. Our findings offer insight to understanding the
intrinsic transport properties of BLGs.

Our DFT calculations adopt the Perdew-Burke-Ernzerhof version of
exchange-correlation functional~\cite{Perdew} and the projector augmented wave
method~\cite{Kresse1} for ionic potentials as implemented in the Vienna Ab
Initio Simulation Package~\cite{Kresse2}.  We obtain interlayer distances
between 3.25 \r{A} (AB) and 3.45 \r{A} (AA) with van der Waals
correction~\cite{Tkatchenko}; interlayer distance of all the configurations is
fixed at 3.35 \r{A} (unless specified) with practically no changes in their band
structures.  To ensure an accurate bandgap, the 2D DFT band structure near the
$K$ point is interpolated~\cite{Waninter} using maximally localized Wannier
functions~\cite{MarzariSouza}.  Effective Hamiltonians are constructed with the
obtained hopping parameters truncated to the first nearest interlayer hoppings
(see details in Supplemental Materials).

\begin{figure}[t]
\includegraphics[width=1.0\columnwidth]{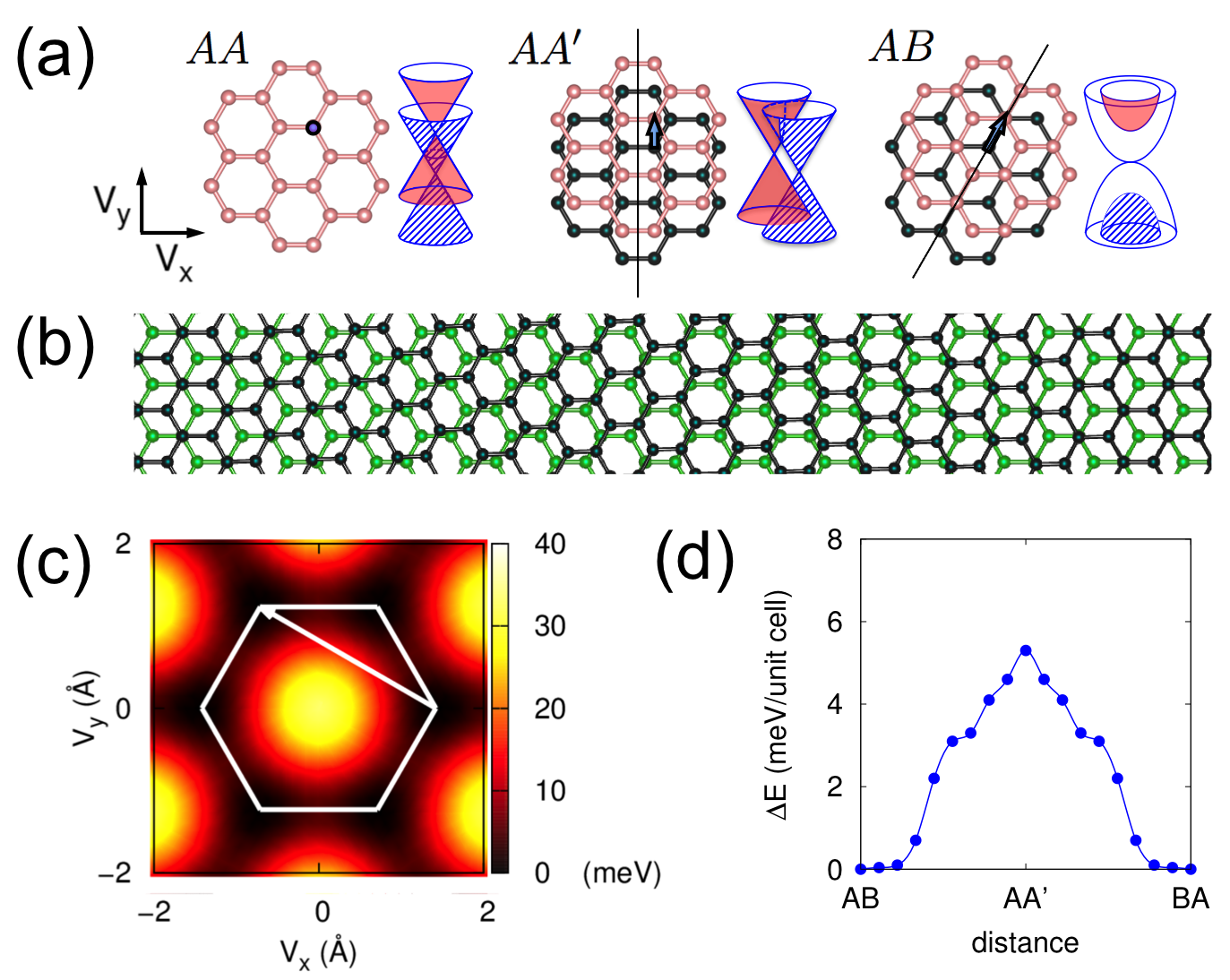}
\caption{(Color online).  (a) Schematic band structures of AA, AA$^\prime$, and
AB.  The solid lines are reflection planes, where the translation vectors
($V_x$,$V_y$) describe the relative displacement between the two layers in the
$xy$ plane. (b) Modeling of the two AB-stacking boundary.  (c)
Stacking-dependent potential energy of BLG per unit cell, where the origin
corresponds to AA stacking. A lattice Wigner-Seitz cell is highlighted by the
solid white line, and the arrow denotes the displacement vector between the two
AB-stacking domains shown in (b). (d) Minimum energy path between the two AB
stackings.
} 
\label{Fig1}
\end{figure}
%
One of the intriguing properties of BLG is that a change in weak interlayer
interaction (which is an order of magnitude smaller than the intralayer coupling
strength) accompanied by a modification in stacking configuration can
significantly alter the electronic structure around the Fermi level.
{\color{black} Figure \ref{Fig1}(a) illustrates schematic band structures of AA,
AA$^\prime$, and AB stacking, where we define systems with equivalent two
sublattices, such as AA and AA$^\prime$, as sublattice-symmetric systems;
otherwise, they are sublattice-asymmetric stackings (such as AB).}  

Figure~\ref{Fig1}(b) shows the atomistic modeling of an experimentally observed
domain boundary~\cite{Alden}.  Here, for visual clarity, we made the phase
boundary region (the region of continuous structural transition between two
AB-stacking regions) much smaller than the experimental ones.  Figure
\ref{Fig1}(c) plots stacking-dependent potential energy with optimized
interlayer distances in the 2D translation vector space, where AB stacking is
used as a reference point.  The arrow denotes the displacement vector between
the left and right domains in Fig.  \ref{Fig1}(b).  Local stacking
configurations of the transition region are distributed on this arrow.  To
remove this soliton-like boundary, one needs to displace the one on the left or
right domain by {\color{black}a displacement vector}.  The minimum energy path
between the two AB stackings lies along the edge of the hexagon with an energy
barrier of 5.3 meV/cell (see Fig.~\ref{Fig1}(d)).  Though this energy barrier
seems quite small, the stacking domain should move as a whole so that the high
energy barrier, which is proportional to the area of the domain ($>10^4$ unit
cell), should be overcome.  This explains the observed stability of non-AB
stacking regions.

The gap-opening mechanism of BLGs can be highly stacking dependent.  Thus, we
first examine the individual band structures near the high-symmetry stackings.
We then discuss the gap-opening properties across complex domain boundaries.
Their effective Hamiltonian in crystal momentum ($k$) space can be described by
a 4 $\times$ 4 matrix with the basis $A_{up}$, $B_{up}$, $A_{dn}$, and $B_{dn}$,
where $A$ and $B$ represent sublattice indices and the subscripts $up$ and $dn$
denote the upper and lower graphene layer, respectively.  Here, the 2 $\times$ 2
block-diagonal components correspond to the individual graphene layers while all
others describe interlayer coupling.  We will now focus only on the effective
Hamiltonian near the $K$ points; band structures around $K^\prime$ can be
obtained by applying time-reversal symmetry to those of $K$.  

First we consider the configurations of AA stacking.  Neglecting small
Bloch phase variations under the Fourier transformations of interlayer coupling,
we find that the Hamiltonian of AA stacking around $K$ becomes 
\begin{equation}
H_0(k)+
\begin{bmatrix}
0 &0 & \widetilde{\gamma}_{AA} & 0 \\
0 & 0 & 0 & \widetilde{\gamma}_{AA} \\
\widetilde{\gamma}_{AA}& 0 & 0 & 0   \\
0 &\widetilde{\gamma}_{AA} & 0 &0
\end{bmatrix},
\end{equation}
with $H_0(k)$ defined as
\begin{equation}
H_0(k)\equiv
\begin{bmatrix}
0 & \hbar v_F k_+ & 0& 0 \\
\hbar v_F k_- & 0 & 0 & 0 \\
0 & 0 & 0 & \hbar v_F k_+   \\
0 & 0 & \hbar v_F k_- &0
\end{bmatrix},
\end{equation}
where the Fermi velocity multiplied by the reduced Planck constant becomes $\hbar
v_F \equiv \frac{\partial E}{\partial k} \sim 5.4$ eV$\cdot$\r{A} and
$k_{\pm}\equiv k_y\pm ik_x $.  {\color{black}$\widetilde{\gamma}_{AA}$
($=-0.34$~eV) is obtained by the Fourier transformation of the interlayer
hopping between $A_{up}$ ($B_{up}$) and $A_{dn}$ ($B_{dn}$),
$\gamma_{AA}$~\cite{FT} .} The hopping parameters between $A_{up}$ ($B_{up}$)
and $B_{dn}$ ($A_{dn}$) become zero because the Bloch phases of three interlayer
nearest neighbors cancel each other at the $K$ point{\color{black}; that is,
$\widetilde{\gamma}_{AB}=0$.} 

By changing our basis to the bonding and antibonding state of each sublattice,
the decoupling of two Dirac cones becomes more transparent as follows: 
\begin{equation}
H_0(k)+
\begin{bmatrix}
\widetilde{\gamma}_{AA} & 0 & 0 & 0 \\
0 &\widetilde{\gamma}_{AA} & 0 & 0  \\
0 & 0 & -\widetilde{\gamma}_{AA} & 0   \\
0 & 0 & 0 & -\widetilde{\gamma}_{AA}
\end{bmatrix},
\end{equation}
which is a block-diagonal Hamiltonian describing two Dirac cones with energy
shift {\color{black}$\pm \widetilde{\gamma}_{AA}$ as shown in the schematic band
diagram of Fig.~\ref{Fig1}(a)}. 

In the AA$^{\prime}$-stacking configuration, one can {\color{black}also}
explicitly illustrate the decoupling of Dirac cones by changing the basis to
interlayer bonding and antibonding of phase-shifted sublattices
[$\frac{1}{\sqrt{2}}A_{up} (B_{up}) \pm \frac{1}{\sqrt{2}}\exp{(\frac{-2\pi
i}{3})}A_{dn} (B_{dn})$]. The Hamiltonian of AA$^{\prime}$ stacking then becomes  
\begin{equation}
H_0(k)+
\begin{bmatrix}
\widetilde{\gamma}_{AA} & -\widetilde{\gamma}_{AB} & 0 & 0 \\
-\widetilde{\gamma}_{AB} & \widetilde{\gamma}_{AA} & 0 & 0  \\
0 & 0 & -\widetilde{\gamma}_{AA} &\widetilde{\gamma}_{AB}   \\
0 & 0 & \widetilde{\gamma}_{AB} & -\widetilde{\gamma}_{AA}
\end{bmatrix},
\end{equation}
where {\color{black}$\widetilde{\gamma}_{AA}=-0.11$}~eV and
{\color{black}$\widetilde{\gamma}_{AB}=-0.22$}~eV, {\color{black}corresponding to
two Dirac cones separated by 0.22 eV in energy with an additional
0.08~\r{A}$^{-1}$ splitting in $k$-space.  Wavefunctions of the decoupled Dirac
cones of both AA and AA$^{\prime}$ stackings have interlayer antibonding and
bonding characteristics, depicted respectively in red (shaded) and blue
(hatched) in Fig. \ref{Fig1}(a). }

\begin{figure*}[t]
\includegraphics[width=2.0\columnwidth]{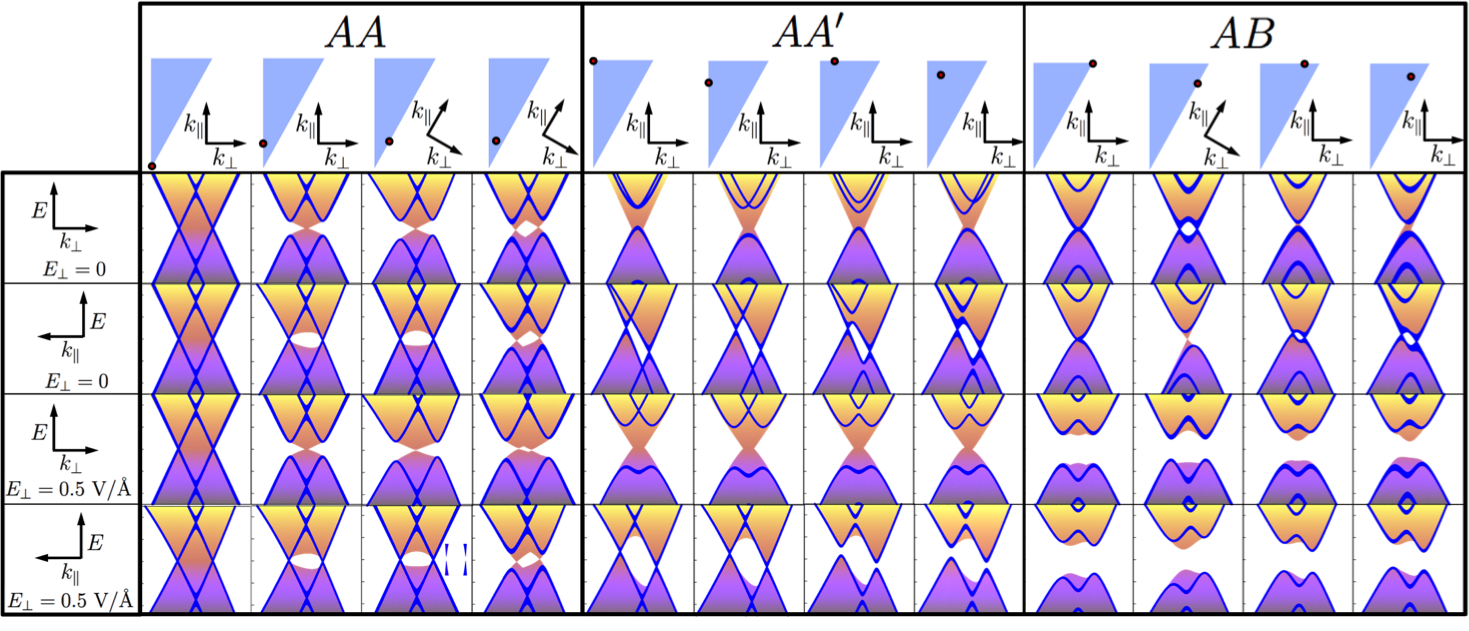}
\caption{(Color online) Projected band structures around $K$ point.
Each configuration is represented by the translation vector in the irreducible
zone of the lattice Wigner-Seitz cell (triangle), where the lower vertex defines
the origin. $k_{\|}$ and $k_{\perp}$ are defined for each configuration. Energy
($k$-space) ranges from $-0.5$~eV ($-0.2$~\r{A}$^{-1}$) to 0.5 eV (0.2
\r{A}$^{-1}$) relative to the Fermi level ($K$). Band structures near the
high-symmetry stackings are projected onto the $k_{\|}$-energy and
$k_{\perp}$-energy planes without and with an E-field. The inset in the third
column of AA stacking highlights a small bandgap ($\approx10$~meV).  
}
\label{Fig2}
\end{figure*}
%
The Hamiltonian of AB staking can be written as 
\begin{equation}
H_0(k)+
\begin{bmatrix}
0 & 0 & 0 & \widetilde{\gamma}_{AB}  \\
0  & 0 & 0 & 0  \\
0 & 0 & 0 & 0  \\
\widetilde{\gamma}_{AB}  & 0 & 0 & 0
\end{bmatrix};
\end{equation}
that is, there are doubly degenerate states at the Fermi level composed of one
sublattice per layer, with no direct coupling between them. In this case two
Dirac points are merged at the $K$ point and split into bonding, antibonding,
and nonbonding types (see Fig.  \ref{Fig1}(a)). 

Next, we trace how a small translation or an external E-field can change the
band structures, especially near the Fermi level.  Our approach is to treat
small translations and E-fields as perturbations to the individual high-symmetry
stackings.

The AA panel of Fig.~\ref{Fig2} summarizes projected (onto $k_{\|}$ and
$k_{\perp}$) band structures for around-AA-stacked graphenes without and with an
E-field.  Since interlayer hopping parameters ($\gamma_{AA}$ and $\gamma_{BB}$)
are the same, one cannot generate an onsite energy difference in the 2 $\times$
2 diagonal block simply by atomic translation, which excludes the direct
coupling between two crossing bands ({\it{i.e.}}, no gap opening).  At the Fermi
level, the hole band of one Dirac cone is degenerated along with the electron
band of the other Dirac cone.  The Fermi surface of the AA stacking is the
intersection of two vertically (energetically) shifted cones, a circle.  A small
translation results in a slight $k$-shift and a coupling of two Dirac cones; a
$k$-shift  changes the circular intersection into a tilted ellipse (which is
still a circle when projected on $k$-space), while a coupling introduces energy
splitting at the intersection. In general, the energy splitting depends on the
angular position of the intersection and becomes zero at two points.  These form
two crossing points near the Fermi level as shown in the second row.  An applied
E-field introduces an additional energy splitting that also depends on the
angular position and becomes zero at two points.  When an E-field is combined
with a sublattice-symmetric translation, their zero splitting points coincide
and the system remains metallic.  In contrast, with the sublattice-asymmetric
translation, each zero coupling points are at a different position and the
crossing points disappear as shown in the inset in the fourth row.  Especially,
when sublattice-asymmetric translation is applied toward AB stacking in the
presence of a reflection and time-reversal symmetry, the minimum bandgaps  occur
along $k_{\|}$ and are located exactly at the same energy.  This means that the
critical field for opening a gap is infinitesimally small.  The perturbational
results on the size of the bandgap are summarized in Table 1. As an example, the
fourth row in the AA block of Fig.  \ref{Fig2} shows a small bandgap of $\sim$10
meV (see the inset) for an asymmetric translation of 0.3 \r{A} and an E-field of
0.5 eV/\r{A}. 

\begin{table*}[t]
\begin{center}
\caption{Analytic expressions of (pseudo-) gaps when a small translation $x$
from reference stacking configurations is combined with an interlayer potential
difference $U$, where $\Delta D_{E(k)}$ denotes energy (crystal momentum)
separation of two Dirac points.}
    \begin{tabular}{ | c | c | c  | c |} 
    \hline
     reference& AA & AA$^\prime$& AB \\ \hline \hline
 {translation} &  \multirow{2}{*}{toward AB} & \multirow{2}{*}{toward AB} & \multirow{2}{*}{any direction} \\ 
    {direction} & & &  \\ \hline
     \multirow{3}{*}{(pseudo-) gap} 
  &   \multirow{3}{*}{$\hbar v_F \frac{\displaystyle \Delta D_k(x)}{\displaystyle 2\Delta D_E(x)} U$} 
   &  \multirow{3}{*}{$\frac{\displaystyle \Delta\widetilde\gamma(x)}{\displaystyle \Delta D_E(x)} U$} 
   &  \multirow{3}{*}{$\frac{\displaystyle\widetilde{\gamma}_{AB}}{\displaystyle \sqrt{\widetilde{\gamma}_{AB}^2+U^2}}U$} \\ 
 & & & \\ 
  &  &  & \\ \hline
    \multirow{3}{*}{parameters} & $\hbar v_F = 5.4$ eV$\cdot$\r{A}  
      & $\Delta\widetilde\gamma(x)\equiv\text{Re}\left(\exp(-\frac{2\pi}{3}i) \left(\frac{\widetilde{\gamma}_{AB}-\widetilde{\gamma}_{BA}}{2} \right)\right) $ & $\widetilde{\gamma}_{AB}=0.30$ eV \\ 
     & $\Delta D_k(x) = 0.03x$ \r{A}$^{-2}$  & $0.2x<\Delta\widetilde\gamma(x)<0.3x $ (eV/\r{A})   &  \\ 
     & $\Delta D_E(x) = 0.68$ eV & $\Delta D_E(x) = 0.22$ eV  &  \\ \hline
     $U$ ($E=0.5$ V/\r{A}) & 0.15 eV & 0.52 eV &0.55 eV\\ \hline
    \end{tabular}
\end{center}
\label{Table1}
\end{table*}
%
Changes in band structures for around-AA$^{\prime}$ stacking are well pronounced
in the $k_{\|}  = 0$ plane [the blues lines in the second and fourth rows of the
AA$^{\prime}$ block in Fig. \ref{Fig2}].  Of the four bands in that plane, only
different Dirac cones can be coupled by a translation. In contrast, under
sublattice-symmetric translation, only parallel-band pairs of each Dirac cone
are coupled, resulting in a balanced repulsion between them.  On the other hand,
under sublattice-asymmetric translation only non-parallel-band pairs of each
cone are coupled, which induces an unfavorable crossing. Also in this slice,
E-field only couples parallel-band pairs for sublattice-symmetric translation.
Though the crossing point in a Dirac cone does not open, each Dirac cone's
crossing band now has a small component of the opposite Dirac cone.  If we apply
an E-field with sublattice-symmetric stacking, crossing bands still remain
crossed because in a Dirac cone, one crossing band does not have a component
parallel to the other crossing band.  But if an E-field is applied to
sublattice-asymmetric stacking, each crossing band now has a small component
parallel to the other crossing bands, which opens a small bandgap.  In spite of
Dirac cones opening, the energy level of each Dirac point is different [the
fourth row in the AA$^{\prime}$ block of Fig.  \ref{Fig2}], thus a relatively
strong E-field is required to change this pseudogap into a true gap.

Finally, we move on to the properties of around-AB stacking.  
As the stacking deviates from exact AB, the doubly degenerate states
of AB at the $K$ point immediately split into two
crossing points. This feature is demonstrated in the second and third columns of
the AB block in Fig.~\ref{Fig2}.  From a symmetry viewpoint, the threefold
rotational symmetry of monolayer graphene is recovered in AA- and AB-stacked
BLG. Combined with translational symmetry, this imposes a threefold symmetry
around the $K$ point. Because two separated crossing points are not compatible
with the symmetry, wavefunction symmetries change during the merging of two
crossing points. It was reported that this merging process is very sensitive to
miniscule translation ($\sim$0.01 \r{A}) and the band topology near the Fermi
level changes~\cite{Son}. Around AB stacking, an E-field opens a
bandgap. Especially, from the eigenvalues of the Hamiltonian, the bandgap
is {\color{black}$\frac{\displaystyle\tilde{\gamma}_{AB}}{\displaystyle
\sqrt{\tilde{\gamma}_{AB}^2+U^2}}U$, where $\tilde{\gamma}_{AB} = \gamma_{AB} =
0.30$ eV.  All the perturbational results for the bandgap are summarized in
Table 1.}

\begin{figure}[t]
\includegraphics[width=1.0\columnwidth]{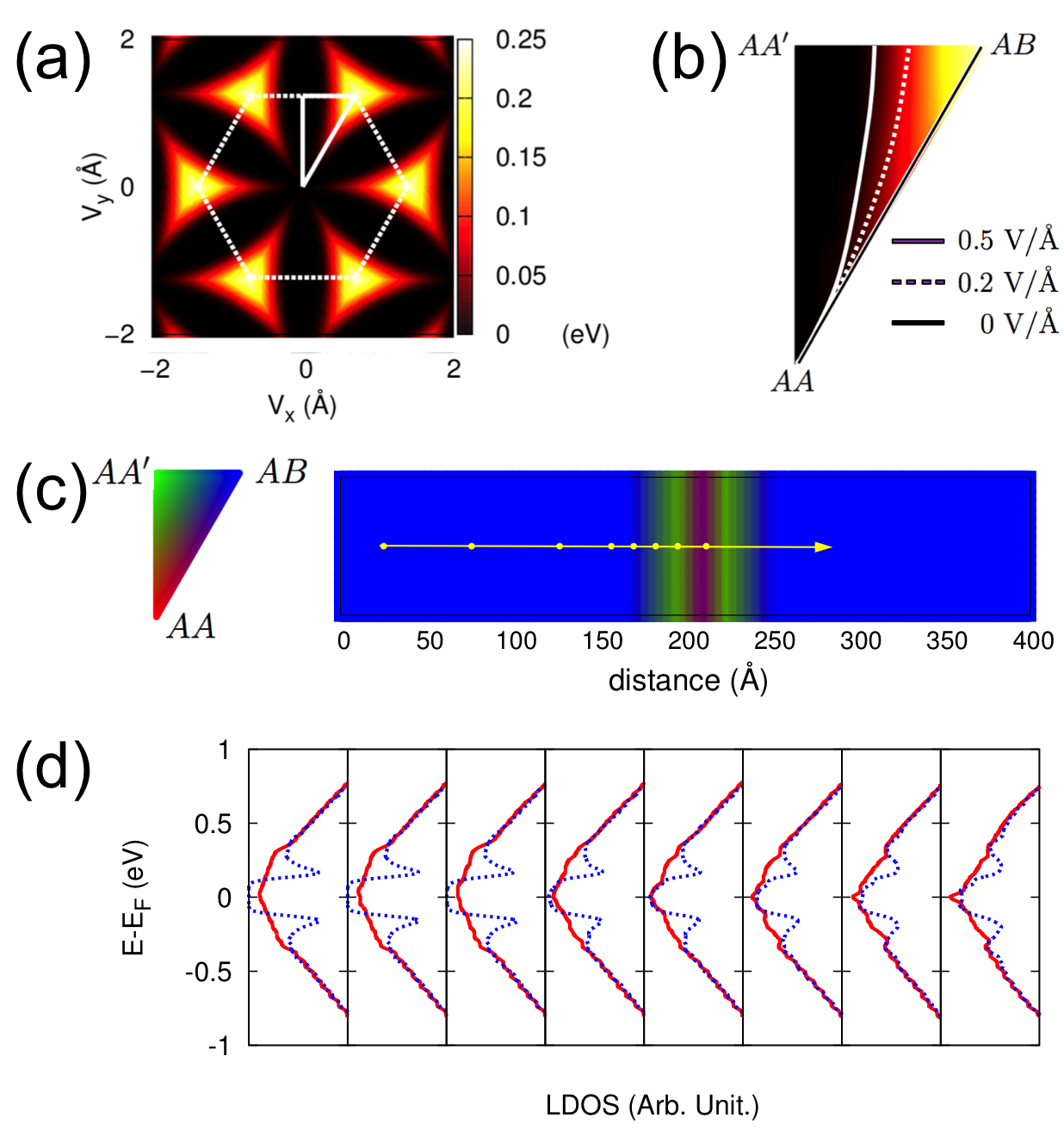}
\caption{(Color online) (a) A stacking-dependent bandgap under a
perpendicular E-field of 0.5 V/\r{A}. A lattice Wigner-Seitz cell is
shown by broken lines and an irreducible zone by solid lines. (b)
Metal-semiconductor phase boundaries for different electric field strengths are
shown for the irreducible zone.  (c) Local stacking configurations of simulated
structure are represented by colors in the triangle at left. (d) Local densities of
states of the spotted region in (c) are plotted from the left with (dotted line) and
without (solid line) an E-field. } \label{Fig3}
\end{figure}
%
Figure~\ref{Fig3}(a) presents the stacking-dependent bandgap under a
perpendicular E-field of 0.5 V/\r{A}. A sizable bandgap opens only around the AB
stacking while the rest still remains metallic. As E-field goes to zero, the
metal-semiconductor phase boundary approaches the line connecting AA and AB
stacking, and the entire region becomes metallic as shown in Fig. \ref{Fig3}(b).
Though no bandgap opens by a pure translation, a minute bandgap ($<$7 meV) was
reported \cite{Shallcross} for a specific rotation angle without any E-field.
To investigate the effect of the non-AB stacking region on the transport
property, we constructed an atomic model of a stacking domain
boundary with a transition length of 50 \r{A} (Fig.  \ref{Fig3} (c)).
Tight-binding parameters are assigned to each atom according to its local
stacking configuration~\cite{TBassign}.  When 0.5 eV of onsite energy difference
between two layers is applied, which corresponds to 0.5 V/\r{A} of E-field, a
bandgap opens at the AB stacking region while there remains finite density of
states at the non-AB stacking region (Fig. \ref{Fig3}(d)).  This indicates that
a high density of midgap states is localized along stacking boundaries even
under a strong E-field.  Because the apparent transport gap is actually
estimated from the activation energy of the carrier, a conduction through these
midgap states can explain the small transport gap and the low-temperature
hopping transport in dual-gated devices. 

In summary, we theoretically investigated stacking-dependent gap-opening
properties of symmetry-broken bilayer graphenes, and established a bandgap phase
diagram.  Our findings may prove to be instrumental in developing graphene-based
electronic devices.

\section*{Acknowledgments}
This research was supported by the Center for Nanophase Materials Sciences, 
Oak Ridge National Laboratory by the Scientific User Facilities Division, Office
of Basic Energy Sciences, U.S. Department of Energy. J. R., S. H., and G. K.
were supported by (1) the Nano∙Material Technology Development Program
(2012M3A7B4049888) through the NRF, funded by the Ministry of Science, ICT and
Future Planning; (2) the Priority Research Center Program (2010-0020207);
and (3) the Basic Science Research Program (2013R1A2009131) through NRF, funded
by the Ministry of Education in Korea.


\begin{thebibliography}{43}
\bibitem{Waninter} We use extremely dense grid points (0.001 \r{A}$^{-1}$ grid
spacing) to ensure an energy resolution up to 5 meV.  Maximally localized
Wannier functions are constructed from the Bloch functions of uniformly-sampled
36 $\times$ 36 $\times$ 1 $k$ points in the Brillouin zone.  In the energy range
of $\pm$2~eV from the Fermi level, first-principles band structures at the
sampled $k$-points are exactly reproduced in this interpolation scheme.
\bibitem{FT} A real space interlayer hopping parameter $\gamma$ is related to
$k$-space one $\widetilde\gamma$ by
\begin{equation}
\widetilde{\gamma}_{lm}
=\sum_{\bold{R}_l,\bold{R}_m}e^{iK\cdot(\bold{R}_m-\bold{R}_l)} 
\gamma_{(l,\bold{R}_l)(m,\bold{R}_m)}, \nonumber
\end{equation}
where $l$, $m$ denotes sublattice indices and $\bold{R}_{l(m)}$ denotes position
vector of $l(m)$ sublattice atom and
$K=(\frac{2\pi}{3a},\frac{2\pi}{\sqrt{3}a})$ and $a$ is C-C bond length.
\bibitem{TBassign} Based on the uniformly sampled 487 stacking configurations,
we extract the exact hopping parameters and truncated them up to
nearest-interlayer pairs. The grid data of hopping parameters is used in the
linear interpolative mapping to each atom with the same local stacking
configuration.  Two hundred k-points along the boundary are sampled for the
density of states calculations.
\end{thebibliography}
\end{document}